\documentclass[cits]{PoS}
\usepackage{graphicx}
\usepackage{rotating}

\newcommand{\kms}{km\,s$^{-1}$}
\newcommand{\Mbh}{$M_{\rm BH}$}
\newcommand{\LLedd}{$L/L_{\rm Edd}$}
\newcommand{\msigma}{M$_{\rm BH}$-$\sigma$}
\newcommand{\msigmaoiii}{M$_{\rm BH}$-$\sigma_{\rm [OIII]}$}

\newcommand{\n}{$n_{\rm e}$}

\title{The Narrow-Line Region of Narrow-Line Seyfert 1 Galaxies}

\ShortTitle{The NLR of NLS1 galaxies}

\author{\speaker{Dawei Xu}\\
        National Astronomical Observatories, Chinese Academy of Sciences,
        20A Datun Road, Beijing 100012, China\\
        E-mail: \email{dwxu@nao.cas.cn}}

\author{S. Komossa\\
        Technische Universitaet Muenchen, Lehrstuhl fuer Physik I,
        James Franck Strasse 1/I, 85748 Garching, Germany;
        Excellence Cluster Universe, TUM, Boltzmannstrasse 2, 85748
        Garching, Germany;
        Max Planck Institut fuer Plasmaphysik, Boltzmannstrasse 2,
        85748 Garching, Germany \\
        E-mail: \email{stefanie.komossa@gmx.de}}

\abstract{
 We have studied the physical properties of a sample of 
 narrow-line Seyfert 1 (NLS1) galaxies, and present a 
 summary of our previous results, and new results. 
 In particular, we have previously shown that 
 (1) the locus of NLS1 galaxies on the \msigmaoiii\ plane 
 does follow the relation of non-active galaxies after 
 removing objects obviously dominated by outflows as evidenced 
 by their [OIII] {\em core} blueshifts. We have (2) 
 identified a number of so-called 'blue outliers' with
 large outflow velocities revealed by their emission-line kinematic shifts.
 We also (3) present new correlations and trends which link black hole mass,
 Eddington ratio and physical parameters of the emission-line regions.}

\FullConference{Narrow-Line Seyfert 1 Galaxies and their place in the Universe\\
         April 4-6, 2011\\
         Milano, Italy}

\begin{document}

\section{Introduction}
 Narrow-line Seyfert 1 (NLS1) galaxies are an exceptional subclass of
 active galactic nuclei (AGNs).  As AGNs with the smallest Balmer
 lines of the broad-line region (BLR) and the strongest FeII
 emission, they are placed at one extreme end of the eigenvector 1
 (EV1) parameter space (e.g., \cite{BOROSON92}).
 Observations and interpretations indicate that NLS1 galaxies
 harbor low-mass black holes accreting at a high rate.  As such,
 they may hold important clues on the nature of black hole
 growth and evolution, and of feeding and feedback
 (see \cite{KOMOSSA08review} for a review on NLS1 galaxies).
 Studying the multi-wavelength continuum and emission-line 
 properties of NLS1 galaxies, the links and correlations 
 between them, and the physics that drive them, is therefore 
 of great interest.

 We have therefore created and analyzed an independent, large, 
 homogeneous sample of NLS1 galaxies plus a comparison sample 
 of broad-line Seyfert 1 (BLS1) galaxies, in order to address 
 the following key topics:
 (1) the locus of NLS1 galaxies on the M-$\sigma$ plane;
 (2) the properties of NLS1 galaxies with extreme outflows;
 (3) the differences in the narrow-line region (NLR) density between NLS1 and
     BLS1 galaxies;
 and (4) the correlations and trends which link the physical properties 
 of our AGN sample, and the physical drivers underlying them.
 Results (1)-(3) have been published before, and we summarize here 
 the salient results, and add some new thoughts. 

 Throughout this paper, a cosmology with $H_{\rm 0}=70$\,\kms\,Mpc$^{-1}$,
 $\Omega_{\rm M}=0.3$ and $\Omega_{\rm \Lambda}=0.7$ is adopted.

\section{The sample}

The NLS1 galaxies which make our sample were selected 
from the catalog of V\'eron-Cetty \& V\'eron \cite{VERON}, to which we 
added a comparison sample of BLS1 galaxies
from \cite{BOROSON03} at $z < 0.3$. The sample was first 
presented by \cite{XU07}.
All galaxies have been observed in the
course of the {\it Sloan Digital Sky Survey} (SDSS)
(Data Release 3; \cite{DR3})
and have detectable low-ionization emission lines (in particular,
[SII]\,$\lambda \lambda 6716,6731$, is always present with S/N $>$ 5).
The initial sample selection, data preparation,
and data analysis methods were described in detail in \cite{XU07}.
Emission lines were fit with Gaussian profiles.
The Balmer lines were decomposed into a narrow and a broad
component, representing emission from the NLR and
BLR, respectively. The broad component itself was fit by
combining two Gaussian profiles.
We adopted the classical FWHM cut-off of the broad component of H$\beta$ of
FWHM(H$\beta_{b}$) $<$ 2000 \kms\ as classification criterion of
NLS1 galaxies. Re-classifying the galaxies of our sample accordingly, 
based on spectral emission-line fitting and FWHM determination, 
we have 55 NLS1 and 39 BLS1 galaxies in our sample.
Emission-line and continuum measurements were then used to derive
AGN parameters, including black hole masses and Eddington ratios.

{\bf Black hole masses: }
we have estimated the black hole masses of
our NLS1 and BLS1 galaxies using  
the width of H$\beta_{b}$ and applying the 
$R_{\rm BLR}$--$\lambda L_{5100}$ relation (\cite{KASPI05}).
The NLS1 galaxies have, on average, smaller black hole masses
than the BLS1 galaxies (Figure\,1a), as commonly found when 
applying these relations.
The estimated black hole masses of NLS1 galaxies range from
$\log\,(M_{\rm BH}/{\rm M}_{\odot})_{\rm NLS1} = 5.7$ to $7.3$
with an average value of $6.5$, while 
BLS1 galaxies show a range from 
$\log\,(M_{\rm BH}/{\rm M}_{\odot})_{\rm BLS1} = 6.5$
to $8.4$ with an average value of $7.2$.

{\bf Eddington ratios:}
Eddington ratios were then derived from the black
hole masses, according to
$L_{\rm Edd} = 1.3\,10^{38}$\,$M_{\rm BH}/M_{\odot}\, {\rm erg\,s}^{-1}$.
The bolometric luminosity $L_{\rm bol}$ was obtained after applying a
correction of $L_{\rm bol}=9\,\lambda{L_{\rm 5100 }}$.
NLS1 galaxies show, on average, higher Eddington ratios than BLS1 galaxies
(Figure\,1b).
Eddington ratios of our NLS1 galaxies range from
$\log\,(L/L_{\rm Edd})_{\rm NLS1} = -0.6$ to $0.3$ 
with an average value of $-0.1$,
while those of BLS1 galaxies range from 
$\log\,(L/L_{\rm Edd})_{\rm BLS1} = -1.4$
to $-0.3$ with an average value of $-0.8$.

%-----------------
\begin{figure}
\begin{center}
\includegraphics[scale=1.0]{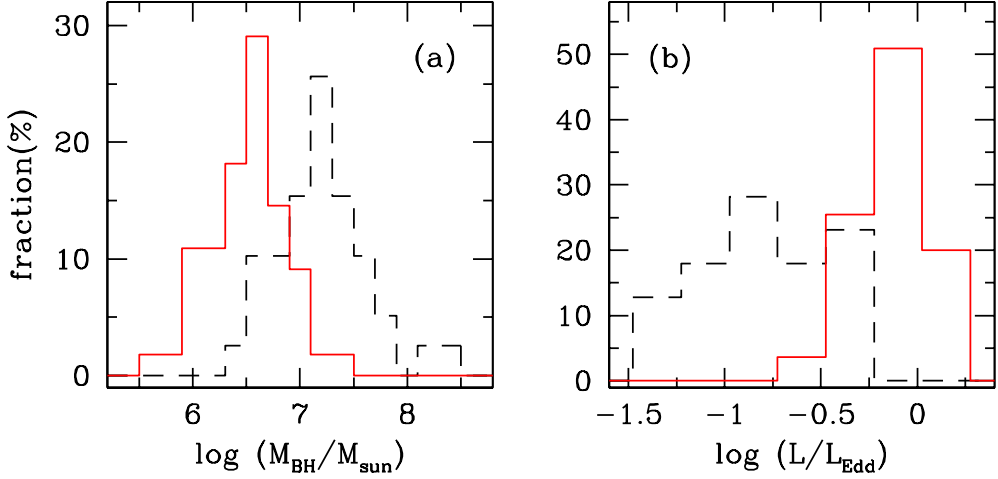}
\caption{Histograms of black hole masses and Eddington ratios
(solid line, NLS1 galaxies; dashed line, BLS1 galaxies).}
\end{center}
\end{figure}
%----------------

\section{The \msigma\ relation}
Studying the locus of NLS1 galaxies on the \msigma\ plane
provides us with important clues on the question,
how galaxies with rapidly growing black holes evolve (\cite{MATHUR01}). 
Previous studies of NLS1 galaxies, employing different
samples and methods, led to partially conflicting results
regarding the galaxies' location
on the \msigma\ plane.
When using the width of the narrow [OIII]\,$\lambda 5007$
emission line, FWHM$_{\rm [OIII]}$, as replacement
for stellar velocity dispersion $\sigma_*$,
some authors found that, on average,
NLS1 galaxies are {\em off} the classical \msigmaoiii\ relation
(e.g., \cite{GRUPE-MATHUR}), while in other studies they turned out to be
{\em on} that relation (e.g., \cite{WANG01}).

We have carefully re-investigated (\cite{KOMOSSA07}) the use of the 
width of [OIII] as a suitable surrogate for $\sigma_*$ 
for the galaxies of our sample.
[OIII]\,$\lambda 5007$ is commonly used for an estimate of $\sigma_*$, but
it is also well-known that this line shows complex and asymmetric profiles (see, e.g., \cite{GRUPE-MATHUR}), and other phenomena (see 
below), occasionally. Special attention was therefore paid to modeling
the [OIII] profiles.
The total [OIII] profile, [OIII]$_{\rm total}$ was decomposed
into two Gaussian components: a narrow core and a broad base.
The narrow core of [OIII] is referred to as [OIII]$_{\rm c}$.
We distinguish between two types of [OIII] spectral complexity:
(1) the presence of a broad base, which tends to be blue-asymmetric
and is referred to as "blue wing"; and (2) systematic blueshifts of the
{\em whole core} of [OIII].

We routinely corrected for blue wings (i.e., we removed them), and only the
narrow core of [OIII] was used to measure $\sigma$. 
Remarkably, a fraction of all NLS1 galaxies show a significant blueshift
of their whole (symmetric) [OIII] {\em core} profile, 
with a velocity shift $v_{\rm [OIII]_c} > 150$ \kms (hereafter referred to
as "blue outliers" \cite{ZAMANOV}; marked in Figure\,2).
This blueshift comes with a dramatic extra {\em broadening} of the 
[OIII] core profile, and it is actually exactly these NLS1 galaxies
which show a significant offset from the classical  \msigmaoiii\ relation.

%----------------
\begin{figure}
\begin{center}
\includegraphics[scale=1.0]{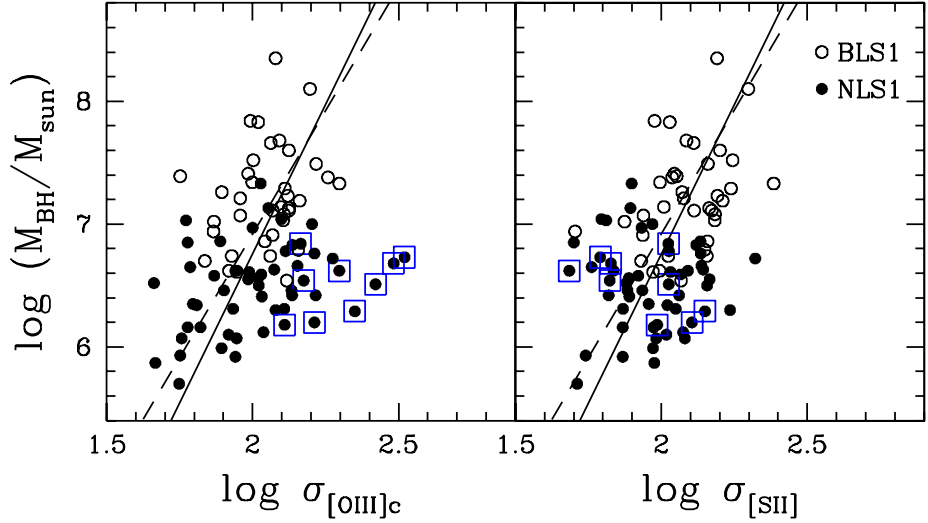}
\caption{Location of NLS1 galaxies (filled circles) and BLS1 galaxies
(open circles) on the \msigma\ plane. Blue outliers in [OIII] are
marked with extra open squares.
{\it Left:} $\sigma$ measurements are based on the narrow core of
[OIII]\,$\lambda 5007$. {\it Right:} $\sigma$ measurements are based on
[SII]. The dashed and solid lines represent the \msigma$_*$ relation
of non-active galaxies of \cite{TREMAINE}
and of \cite{FERRARESE}.}
\end{center}
\end{figure}
%----------------

Since the velocity fields of these blue outliers are not dominated
by the bulge potential, their line widths cannot be used as
surrogates for $\sigma_*$. In our sample, all remaining NLS1 
galaxies are {\it on} the \msigmaoiii\ relation,
after excluding the objects with strong [OIII] core blueshifts (Figure\,2).
The full results of this study have been presented by \cite{KOMOSSA07}. 

In the future, it will be useful to check, whether other NLS1 samples also
contain [OIII] blue outliers, which have their whole line cores blueshifted
(we would like to re-emphasize here, 
that our key finding is about the core blueshifts,
not about the blue wings).  

Furthermore, we have also explored the usefulness 
of [SII]\,$\lambda \lambda 6716,6731$
as a surrogate for $\sigma_*$.
We found that NLS1 galaxies do follow the same \msigma\ relation
as BLS1 galaxies if the width of [SII] is used as a 
substitute for $\sigma_*$.
We have checked that [SII] is {\em not} systematically influenced
by the outflow component that appears in [OIII]$_{\rm c}$.
NLS1 galaxies with blue outliers in [OIII]$_{\rm c}$ are still
distributed along the classical $M_{\rm BH} - \sigma$ relation 
when using [SII] (Figure\,2).

Zhou et al. \cite{ZHOU} presented estimates of stellar velocity
dispersion from SDSS spectra, and they concluded that NLS1
galaxies were highly offset from the classical the \msigma$_*$ relation 
(but see \cite{BOTTE} who found that NLS1 galaxies were on the relation 
based on the measurements of stellar absorption lines).

Can these different results be reconciled with each other
in one consistent model? That is to say, can we account for
the fact that some estimates of $\sigma_*$ place NLS1 galaxies
to the right of the classical relation in \msigma\ diagrams, 
while our emission-line measurements put them on the relation? 
There is strong evidence, that a significant fraction of barred 
galaxies does not actually follow the classical \msigma$_*$ relation, 
but is rather offset (e.g., \cite{HU}) .
The same may therefore hold for barred NLS1 galaxies. 
Since the stars follow the bar, the possibility has been discussed, 
that streaming motions along the bar, or other effects, influence
the measurements of $\sigma_*$. On the other hand, the NLR may 
not follow the bar, but may rather be confined to the central core, 
and is always unresolved in our sources;
i.e., we always include the whole NLR in our measurements. 
Therefore, velocity dispersion measurements of stars and gas 
in barred galaxies may lead to different results. 

%----------------
\begin{figure}
\begin{center}
\includegraphics[scale=1.0]{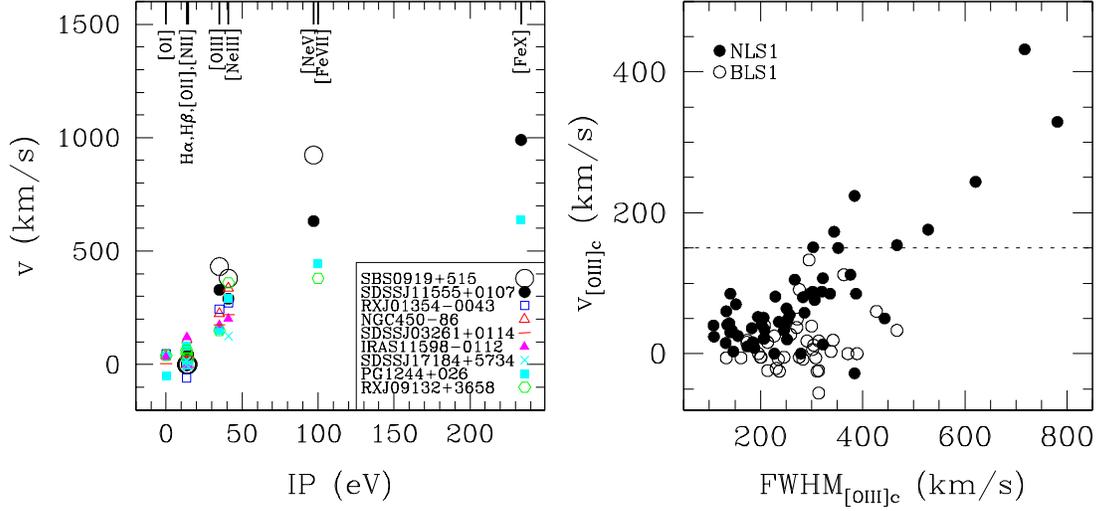}
\caption{{\it Left:} velocity shift of individual emission lines vs.
the ionization potential IP of [OIII] blue outliers.
{\it Right:} correlation of the [OIII]
core width with the [OIII] core blueshift.}
\end{center}
\end{figure}
%-------------

\section{Blue outliers}

Among our sample, we identified in total nine NLS1 galaxies 
as "blue outliers",  as described above.
In a follow-up study, we have systematically explored
the optical properties (\cite{KOMOSSA08blueoutliers})
of these blue outliers. Our main results can be summarized as follows: 
(1) All of them show high Eddington ratios and narrow Balmer 
lines of the BLR.
(2) We detected a strong correlation of line blueshift with
ionization potential in each galaxy, and confirmed a strong
correlation between [OIII] blueshift and line width (Figure\,3). 
(3) We also reported the absence of a zero-blueshift [OIII] component
from a classical inner NLR, while the presence of a
classical quiescent outer NLR is indicated by the existence
of low-ionization lines.

The high Eddington ratios possibly lead to strong radiation-pressure
driven winds/outflows. Blue outliers with their
extreme velocity shifts place tight constraints on models of
the NLR and mechanisms of AGN outflows on large scales.
We tentatively favor a scenario where the NLR clouds of blue outliers
are entrained in a decelerating wind 
(see \cite{KOMOSSA08blueoutliers} for more details).

\section{The NLR density}

When attempting to explain the multi-wavelength properties of NLS1 galaxies,
and correlations among them, winds and density effects have 
been suggested as potentially important secondary parameters, 
driving some of the observed trends. We have therefore made use 
of the diagnostic power of the [SII]\,$\lambda \lambda6716,6731$
intensity ratio to measure the density of the NLR, and have 
investigated whether or not there is any difference in the 
NLR density between NLS1 galaxies and BLS1 galaxies. We found 
that the average NLR density of NLS1 galaxies is
lower than that of BLS1 galaxies. There is a '{\em zone of avoidance}'
(\cite{XU07}) in density in the sense that BLS1 galaxies avoid low densities,
while NLS1 galaxies show a larger scatter in densities
including a significant number of objects with low densities (Figure\,4).

%---------------
\begin{figure}[!t]
\begin{center}
\includegraphics[scale=0.8]{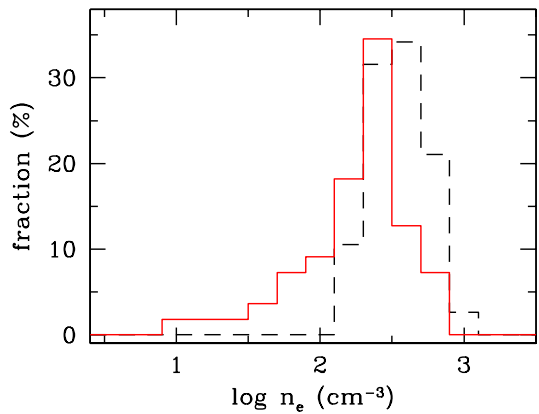}
\caption{Histograms of the NLR density
(solid line, NLS1 galaxies; dashed line, BLS1 galaxies).}
\end{center}
\end{figure}
%---------------

We found several lines of evidence that outflows
play a significant role in driving the difference in the NLR
between NLS1 and BLS1 galaxies.
If the radiation-pressure driven winds/outflows still propagate
up into the NLR, then we may expect that the NLR gas in such objects
is actually more tenuous. An anti-correlation of decreasing electron
density with increasing Eddington ratio can be seen across our entire
sample of NLS1 and BLS1 galaxies. We found that the peak blueshift 
of [OIII] does significantly correlate with \LLedd.
This correlation then indicates that outflows are more common
in objects with high accretion rates.
We tentatively favor the effects of winds/outflows, stronger
in NLS1 galaxies than in BLS1 galaxies, as explanation
for the zone of avoidance in density (see
\cite{XU07} for full results).

\section{Correlation analysis}

In order to study trends across our sample,
and to make a comparison with previously known correlations
derived for different NLS1 samples, and to identify new trends,
we have performed two types of correlation analyses.
Independent samples are of importance when assessing the robustness
of correlation analysis, as is the need to increase correlation space.
Ours is a large, homogeneously analyzed sample, and we add new
emission-line measurements (particularly, the
the density-sensitive [SII] ratio) to correlation analysis.

In a first step we performed a Spearman rank order correlation analysis
between the key parameters measured for our sample.
These correlations involve emission-line widths
(FWHM$_{\rm H\beta_b}$, FWHM${\rm_{[SII]}}$ and FWHM${\rm_{[OIII]_c}}$),
emission-line ratios (the strength of the [OIII] line, abbreviated as R5007,
the strength FeII complex R4570 and 
the sulphur emission-line ratio R${\rm_{[SII]}}$
defined as the intensity ratio of [SII]\,$\lambda 6716/\lambda 6731$,
and the inferred NLR density \n),
and the parameters $\lambda{L_{\rm 5100}}$, \LLedd, and \Mbh.
Figure\,5 displays the correlation diagrams for several parameters.
The full Spearman rank order correlation analysis
will be presented in \cite{XU11}.

%--------------
\begin{figure}
\begin{center}
\includegraphics[scale=0.6]{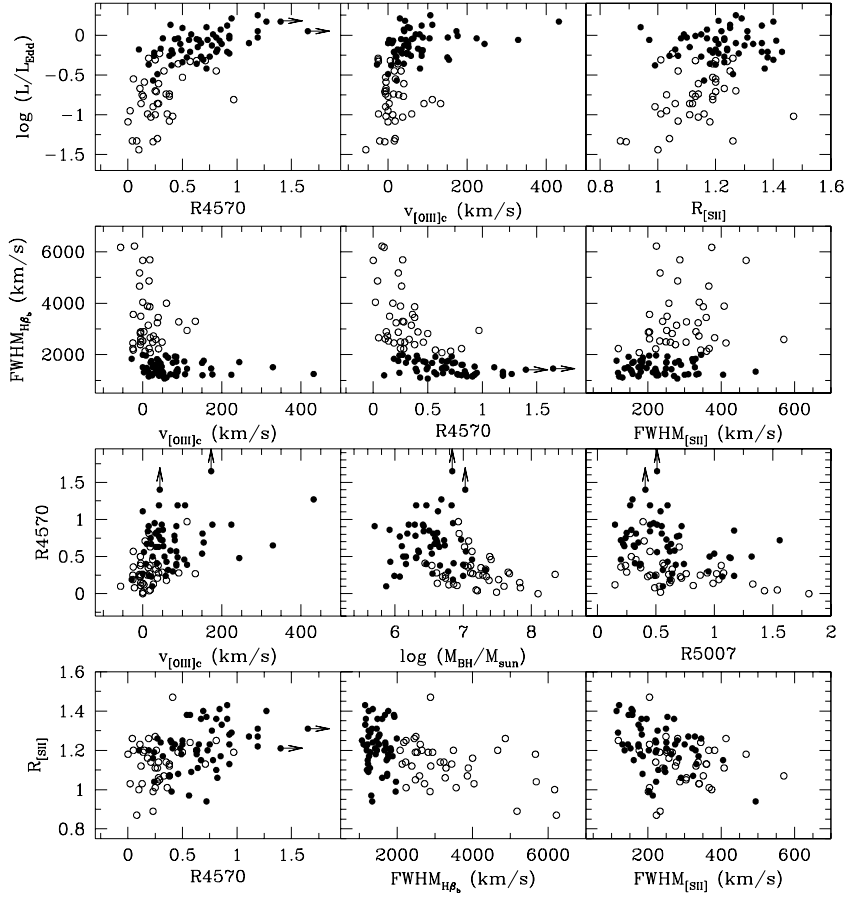}
\caption{
Strong correlations among
black hole mass, Eddington ratio and emission-line properties.
Symbols are as in Figure\,2. 
The objects that are off the plots are indicated by arrows.}
\end{center}
\end{figure}
%--------------

We then performed a Principle Component Analysis (PCA;
\cite{BOROSON92,FRANCIS}), in order to identify the main drivers of the
correlation properties. We involved the NLR density (represented
by the Sulphur ratio R${\rm_{[SII]}}$) in the PCA for the first time.
We found EV1 is significantly (anti-)correlated with R4570,
FWHM$_{\rm H\beta_b}$,
R${\rm_{[SII]}}$, FWHM${\rm_{[SII]}}$ and $v_{\rm_{[OIII]_c}}$.
When EV1 decreases, R4570 increases, outflow
velocity becomes stronger and R${\rm_{[SII]}}$ increases,
while R5007, FWHM$_{\rm H\beta_b}$ and FWHM${\rm_{[SII]}}$ decrease.
A correlation analysis further shows that EV1 strongly correlates
with \LLedd\ and the NLR density \n\ (Figure\,6).
We confirm that EV1 is an indicator of \LLedd\ 
(e.g.,\cite{BOROSON92,SULENTIC02, XU03,GRUPE11}. 
We further identify the NLR density as a new EV1 element for AGNs.

%------------------
\begin{figure}
\begin{center}
\includegraphics[scale=0.9]{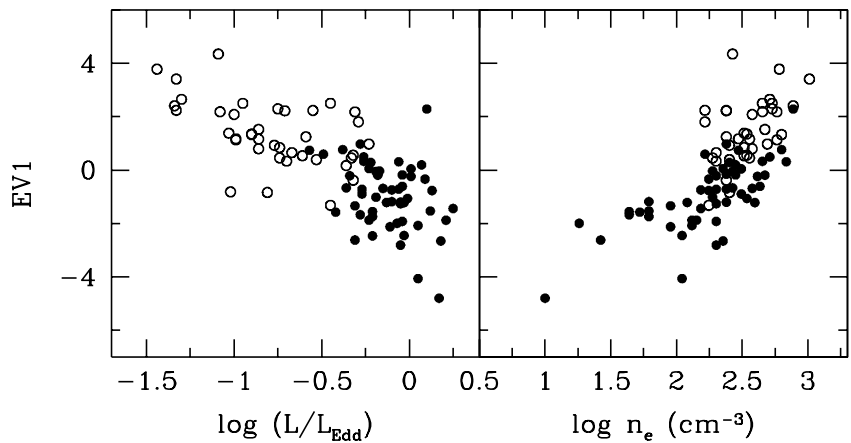}
\caption{Correlations of EV1 with $L/L_{\rm Edd}$ and
$n_{\rm e}$. Symbols are as in Figure\,2. The EV1 is
plotted in arbitrary units.}
\end{center}
\end{figure}
%------------------

\section{Conclusions}

We have systematically studied a large sample of
NLS1 galaxies plus a comparison sample of BLS1 galaxies, and obtained
the following results:

\begin{itemize}
\item NLS1 galaxies in our sample do follow the \msigma\ relation
of BLS1 galaxies and non-active galaxies, if [SII], and [OIII]$_c$ are used
to measure $\sigma$; {\rm with one exception:}
galaxies with systematic blueshifts of the {\em whole core profile of} 
their [OIII] line also have anomalously broad [OIII] profiles. 
The velocity fields of these "blue outliers" are dominated by
outflows and are therefore not suitable for $\sigma_*$ measurements. All
samples making use of [OIII] should therefore have blue outliers removed, 
before measuring velocity dispersion from [OIII].

\item The [OIII] blue outliers are of independent interest because of
their strong large-scale outflows, which also appear in other 
high-ionization emission lines. 

\item NLS1 galaxies show lower average NLR density than BLS1 galaxies.
We identify the NLR density as a new element of EV1 in AGNs.

\end{itemize}

\acknowledgments
This work is supported by the National Natural Science Foundation
of China (Grant No. 10873017) and National Basic Research Program
of China - 973 program (Grant No. 2009CB824800). S.K. acknowledges the
Aspen Center for Physics for its hospitality.


\begin{thebibliography}{99}

\bibitem{DR3} K. Abazajian, et al.: 
\emph{The Third Data Release of the Sloan Digital Sky Survey},
\emph{AJ} \textbf {129} (2005) 1755.

\bibitem{BOROSON92} T.~A. Boroson \& R.~F. Green:
\emph{The emission-line properties of low-redshift quasi-stellar objects},
\emph{ApJS} \textbf{80} (1992) 109.

\bibitem{BOROSON03} T.~A. Boroson: 
\emph{Does the narrow [OIII]\,$\lambda 5007$ line reflect the stellar 
velocity dispersion in active galactic nuclei?},
\emph{ApJ} \textbf{585} (2003) 647.

\bibitem{BOTTE} V. Botte, et al.: 
\emph{Exploring narrow-line Seyfert 1 galaxies through the physical 
properties of their hosts},
\emph{AJ} \textbf{127} (2004) 3168.

\bibitem{FERRARESE} L. Ferrarese, \& H. Ford: 
\emph{Supermassive black holes in galactic nuclei: past, present 
and future research},
\emph{Space Sci. Rev.} \textbf{116} (2005) 523.

\bibitem{FRANCIS} P.~J. Francis, \& B.~J. Wills: 
\emph{Introduction to Principal Components Analysis}, In:
\emph{ASP Conf. Series} \textbf{162} (1999) 363.

\bibitem{GRUPE-MATHUR} D. Grupe, \& S. Mathur:  
\emph{\msigma\ relation for a complete sample of soft X-ray-selected 
active galactic nuclei},
\emph{ApJ} \textbf{606} (2004) 41.

\bibitem{GRUPE11} D. Grupe:
\emph{Statistical analysis of an AGN sample with simultaneous UV and 
X-ray observations with Swift}, In:
\emph{Proceedings of the Workshop Narrow-Line Seyfert 1 Galaxies 
and Their Place in the Universe}, \pos{PoS(NLS1)004} (2011).

\bibitem{HU} J. Hu: 
\emph{The black hole mass-stellar velocity dispersion correlation: 
bulges versus pseudo-bulges},
\emph{MNRAS} \textbf{386} (2008) 2242.

\bibitem{KASPI05} S. Kaspi, et al.: 
\emph{The relationship between luminosity and broad-line region 
size in active galactic nuclei},
\emph{ApJ} \textbf{629} (2005) 61.

\bibitem{KOMOSSA07} S. Komossa, \& D. Xu: 
\emph{Narrow-line Seyfert 1 galaxies and the \msigma\ relation},
\emph{ApJ} \textbf{667} (2007) L33.

\bibitem{KOMOSSA08review} S. Komossa: 
\emph{Narrow-line Seyfert 1 galaxies},
\emph{RevMexAA (Serie de Conferencecias)} \textbf{32} (2008) 86.

\bibitem{KOMOSSA08blueoutliers} S. Komossa, et al.: 
\emph{On the nature of Seyfert galaxies with high [OIII]\,$\lambda 5007$ 
blueshifts},
\emph{ApJ} \textbf{680} (2008) 926.

\bibitem{MATHUR01} S. Mathur et al.:
\emph{Evolution of active galaxies: black-hole mass-bulge relations 
for narrow line objects},
\emph{New Astron.} \textbf{6} (2001) 321.

\bibitem{SULENTIC02} J.~W. Sulentic et al.: 
\emph{Average quasar spectra in the context of eigenvector 1},
\emph{ApJ} \textbf{566} (2002) L71.

\bibitem{TREMAINE} S. Tremaine, et al.: 
\emph{The slope of the black hole mass versus velocity 
dispersion correlation},
\emph{ApJ} \textbf{574} (2002) 740.

\bibitem{VERON} M.~P. V\'eron-Cetty \& P. V\'eron: 
\emph{A catalogue of quasars and active nuclei: 11th edition},
\emph{A\&A} \textbf{412} (2003) 399.

\bibitem{WANG01} T. Wang, \& Y. Lu: 
\emph{Black hole mass and velocity dispersion of narrow line region in active galactic nuclei and narrow line Seyfert 1 galaxies},
\emph{A\&A} \textbf{377} (2001) 52.

\bibitem{XU03} D. Xu, et al.:
\emph{An active galactic nucleus sample with high X-Ray-to-optical 
flux ratio from RASS. II. Optical emission line properties of 
Seyfert 1-type active galactic nuclei},
\emph{ApJ} \textbf{590} (2003) 73.

\bibitem{XU07} D. Xu, et al.:
\emph{The narrow-line region of narrow-line and broad-line type 1 
active galactic nuclei. I. A zone of avoidance in density},
  \emph{ApJ} \textbf{670} (2007) 60.

\bibitem{XU11} D. Xu, et al.: 2011, in preparation

\bibitem{ZHOU} H. Zhou, et al.: 
\emph{A comprehensive study of 2000 narrow line Seyfert 1 galaxies 
from the Sloan Digital Sky Survey. I. The sample},
\emph{ApJS} \textbf{166} (2006) 128.

\bibitem{ZAMANOV} R. Zamanov, et al.: 
\emph{Kinematic linkage between the broad- and narrow-line-emitting gas 
in active galactic nuclei}, 
\emph{ApJ} \textbf{576} (2002) L9.

\end{thebibliography}
\end{document}